\documentclass[useamsfonts]{pasj00}

\usepackage{graphicx}
\usepackage{natbib}

\newcommand{\bx}{\bar{x}}
\newcommand{\by}{\bar{y}}
\newcommand{\bsigma}{\bar{\sigma}}
\newcommand{\bomega}{\bar{\omega}}
\newcommand{\bkappa}{\bar{\kappa}}

\newcommand{\dd}{\mathrm{d}}
\newcommand{\vc}[1]{\mbox{\protect\boldmath$#1$}}              

\newcommand{\ii}{\mathrm{i}}
\newcommand{\Fv}{\mathcal{F}_\mathrm{visc}}

\newcommand{\ie}{i.e.\ }

\usepackage{color}

\begin{document}

\SetRunningHead{J.~Hor\'ak et al.}{Oscillations of viscous slender tori}
\Received{2011 .... ....}  
\Accepted{2011 .... ....}

\title{Alpha-viscosity effects in slender tori}

\author{
  J.~\textsc{Hor\'ak},\altaffilmark{1}
  M.\,A.~\textsc{Abramowicz},\altaffilmark{2,4,5}
  L.~\textsc{Levin},\altaffilmark{2,3}
  R.~\textsc{Slapak}\altaffilmark{2}
  and
  O.~\textsc{Straub}\altaffilmark{4}
}

\affil{$^1$Astronomical Institute, Academy of Sciences,
   Bo\v{c}n\'{\i}~II, CZ-141\,31 Prague, Czech Republic}
\affil{$^2$Department Physics, G{\"o}teborg University,
   S-412\,96 G{\"o}teborg, Sweden}
\affil{$^3$Swinburne University of Technology, 
   Centre for Astrophysics and Supercomputing, \\
   Mail H30, PO Box 218, VIC 3122, Australia}
\affil{$^4$Copernicus Astronomical Centre PAN, Bartycka 18, 
   00-716 Warszawa, Poland}
\affil{$^5$Institute of Physics, Faculty of Philosophy and Science, 
   Silesian University in Opava, \\
   Bezru\v{c}ovo n\'{a}m. 13, CZ-746-01 Opava, Czech Republic}


\maketitle


\begin{abstract}
We explore effects of the Shakura-Sunyaev $\alpha$-viscosity on the dynamics and oscillations of slender tori. We start with a slow secular evolution of the torus. We show that the angular-momentum profile approach the Keplerian one on the timescale longer than a dynamical one by a factor of the order of $1/\alpha$. Then we focus our attention on the oscillations of the torus. We discuss effects of various angular momentum distributions. Using a perturbation theory, we have found a rather general result that the high-order acoustic modes are damped by the viscosity, while the high-order inertial modes are enhanced. We calculate a viscous growth rates for the lowest-order modes and show that already lowest-order  inertial mode is unstable for less steep angular momentum profiles or very close to the central gravitating object. 
\end{abstract}

\section{Introduction}
\label{intro}


Stability and propagation of waves in viscous accretion disks have been studied extensively for more then thirty years, since the pioneering work of \citet{Kato1978}. Adopting the local approximation, he has shown that the horizontal $p$-mode oscillations of the Keplerian isothermal disk become over-stable when a small viscosity of the fluid is introduced. The instability arises because the azimuthal component of the viscous force varies in phase with the azimuthal velocity during oscillations and thus a positive work on the oscillations is done. The source of the energy for the oscillations is the angular momentum flowing radially in the background flow. Under suitable conditions, given mainly by the temperature and density dependence of viscosity and also by the mode eigenfrequencies and the shape of the modal eigenfunctions, this process may provide energy sufficient to amplify the oscillations against other normal viscous damping processes. The instability of $p$-modes is even stronger in the relativistic disk, where the gradient of the angular velocity is steeper.

Later on, this problem has been reconsidered by other authors and the analysis has been expanded also to other types of oscillation modes. For example, \citet{Blumenthal+1984} studied stability of $p$-modes in disks with different ratios of the gas to the radiation pressure, \citet{Kato1991} discussed possibility of viscous excitation of the one-armed corrugation mode in relativistic disks. The local approximation was released in the work of \citet{Nowak-Wagoner1992}, who estimated growth rates of $p$-modes trapped near the inner edge of the disk and $g$-modes trapped near the radius of the maximum of the radial epicyclic frequency.

Some of the analysis go even beyond the standard \citet{Shakura-Sunyaev1973} prescription that is based on the mixing-length theory. \citet{Nowak-Wagoner1992} considered the effects of an anisotropic viscosity and showed that the growth of the trapped $g$-modes is enhanced while that of the $p$-modes is suppressed. Perhaps even more advanced is the analysis of $c$-mode oscillations in relativistic disks by \citet{Kato1994} who uses the transport equations in calculating the stress tensor perturbations based on his second-order closure scheme.

In this note we study effects of viscosity on oscillation modes of a slender torus. The main difference of this configuration from the thin disks studied in the work cited above are the angular momentum profiles that can be  significantly different from the Keplerian distribution and the following substantial pressure and density gradient in the radial direction (i.e.\ in the same direction as the angular velocity is changing). The plan of the paper is as follows. In section \ref{sec:inviscid-tori} a basic theory of inviscid tori and the limit of slender tori are reminded. The section \ref{sec:secular-evolution} deals with the secular evolution of the torus that arises when a small viscosity is introduced to the flow. The section \ref{sec:pert} is devoted to the general discussion of oscillation modes and the effects of viscosity on the oscillation frequencies. There we derive a general formula for the eigenfrequency shifts which we subsequently apply in sections \ref{sec:results-special} and \ref{sec:results}. The section \ref{sec:results-special} contains results for the modes of high order and for Keplerian tori and the section \ref{sec:results} focuses on the lowest-order modes. Finally, section \ref{sec:concl} is devoted to discussion and conclusions.

\section{Inviscid slender tori}
\label{sec:inviscid-tori}
The construction of an inviscid axisymmetric polytropic slender torus is described by \citet{Blaes1985} in the Newtonian gravitational field and by \citet{Blaes+2007} in the case of a general axisymmetric gravitational potential. The fluid of the torus is in pure rotation, e.g. its velocity in the polar coordinates $\{r,\phi,z\}$ can be expressed as $v^i=\Omega\,\delta^i_\phi$, where $\Omega$ is the angular velocity. The shape of the equi-pressure and equi-density surfaces coincide because of the polytropic equation of state, $p\propto\rho^{1+1/n}$, and can be uniformly described by the Lane-Embden function $f(\vc{x})$. The density and pressure are then given by $\rho = \rho_0 f^n(\vc{x})$ and $p=p_0 f^{n+1}(\vc{x})$ (the subscript zero indicates the evaluation at the center of the torus, where the density and pressure are maximal). Another consequence of using the polytropic equation of state is that the angular velocity is a function of the radial coordinate only, $\Omega=\Omega(r)$. In addition, throughout the paper, we assume that this function is slowly varying, \ie $\partial_r\Omega\sim\Omega/r$. The structure of these surfaces follows from the poloidal part of the Euler equation
\begin{equation}
  \nabla^i\Phi - r\Omega^2\delta^i_r = -\frac{1}{\rho}\nabla^i p = -\frac{1}{2}\beta^2 r_0^2\Omega_0^2\nabla^i f,
  \label{eq:euler-poloidal}
\end{equation}
where $\beta^2=2(n+1)p_0/(\rho_0 r_0^2\Omega_0^2)$ is the slenderness parameter $\beta$ that gives the thickness of the torus, both the radial and vertical extend of the torus are of the order of $\beta r_0$. 

The limit $\beta\rightarrow0$ describes slender tori. The surfaces of constant pressure and density have elliptic cross-sections, whose centers correspond to the maximal-pressure circle at $r=r_0$ in the equatorial plane. It is convenient to introduce dimensionless coordinates $\bx$ and $\by$ contracting with the torus as $\beta\rightarrow0$,
\begin{equation}
  \bx=\frac{r-r_0}{\beta r_0},
  \quad
  \by=\frac{z}{\beta r_0}.
\end{equation}
The shapes of the equipressure surfaces are then given by
\begin{equation}
  f(\bx,\by) = 1 - (\bomega_r^2 - \bkappa^2)\bx^2 - \bomega_z^2\by^2,
  \label{eq:f}
\end{equation}
where $\bomega_r$ and $\bomega_z$ are the radial and vertical epicyclic frequencies normalized by $\Omega_0$ and $\bkappa^2 = (\dd\ln\ell^2/\dd \ln r)_0$ describes distribution of the angular momentum $\ell=r^2\Omega(r)$, which is in the slender-torus limit linear. It follows from equation (\ref{eq:euler-poloidal}) that $\Omega_0 = \Omega_\mathrm{K}(r_0)$.  The outermost ellipse given by $f=0$ describes the surface of the torus. On the other hand, the torus center at $(\bx,\by)=(0,0)$ corresponds to $f=1$. 

\section{Secular evolution of viscous tori}
\label{sec:secular-evolution}

Introduction of a viscosity causes a slow secular evolution of the torus governed by the Navier-Stokes equation 
\begin{equation}
  \frac{\partial v^i}{\partial t} + v^k\nabla_k v^i + \frac{1}{\rho}\nabla^i p + 
  \nabla^i\Phi = \frac{1}{\rho}\nabla_k\sigma^{ki} = \Fv^i
  \label{eq:NavierStokes}
\end{equation}
The stress tensor of the viscous flow is given by
\begin{equation}
  \sigma^{ik} = \eta\left(\nabla^k v^i + \nabla^i v^k\right) + 
  \left(\xi-\frac{2}{3}\eta\right)(\nabla\cdot\vc{v})g^{ik}, 
\end{equation}
where $\eta$ and $\xi$ are coefficients of dynamic and bulk viscosity, respectively, and $g^{ik}$ is a metric tensor. We extend the standard `$\alpha p$'-parameterization of the shear viscosity $\eta$ traditional in the accretion theory to include also the bulk viscosity $\xi$ by introducing a coefficient $a$,
\begin{equation}
  \eta = \frac{\alpha p}{\Omega_0}, 
  \quad
  \left(\xi-\frac{2}{3}\eta\right) = a\,\frac{\alpha p}{\Omega_0},
  \quad
  \alpha\ll 1.
\end{equation}

The velocity of the fluid is still dominated by the azimuthal component, but also small poloidal components appear due to the viscosity.
In our analysis, we take $\alpha$ as a small parameter with respect to which both the equations and their solutions are expanded. For the poloidal velocity we assume  that $v^i/(r\Omega) = \mathcal{O}(\alpha)$, $i = r,z$, the azimuthal velocity is of the order of the Keplerian orbital velocity at the center of the torus. From a balance among leading-order terms in the poloidal part ($i = r,z$) of the Navier-Stokes equation we obtain again
\begin{equation}
  \frac{1}{\rho}\nabla^i p + \nabla^i\Phi - r\Omega^2\delta^i_r = 0, 
  \label{eq:ns-poloidal}
\end{equation}
while the toroidal part and the continuity equation gives
\begin{equation}
  \frac{\partial\ell}{\partial t}  = - v^r\frac{\partial \ell}{\partial r} + \frac{\alpha}{r\rho} 
  \frac{\partial}{\partial r} \left(r^2 p \frac{d\ln\Omega}{d\ln r}\right)
  \label{eq:ns-toroidal}
\end{equation}
and
\begin{equation}
  \frac{\partial\rho}{\partial t} = - \frac{1}{r}\frac{\partial}{\partial r}(r\rho v^r) - \frac{\partial}{\partial z}(\rho v^z) 
  \label{eq:viscevol-continuity}
\end{equation}
The right-hand sides of the latter two equations are of the first order in $\alpha$, these equations therefore describe a slow evolution of the angular momentum and the density on the time scale $\sim 1/(\alpha\Omega)$. On the other hand, the equation (\ref{eq:ns-poloidal}) states the equilibrium among the pressure, gravitational and centrifugal forces that is reached on much shorter timescale $\sim 1/\Omega$. This equation determines the structure of the equipressure surfaces in the torus and is identical to the case of the inviscid flow. The overall evolution can be therefore regarded as a sequence of inviscid tori whose shapes are given by Lane-Embden function $f$ defined in equation (\ref{eq:f}), but whose main  determining parameters $\bar{\kappa}$, $\beta$ and $r_0$ are slowly changing in time.

Calculating the total mass $\mathcal{M}$ and angular momentum $\mathcal{L}$ of infinitely slender torus by integrating $\rho$ and $\rho\ell$ over the torus volume, we obtain the relation
\begin{equation}
  \mathcal{L} = \ell_\mathrm{K}(r_0) \mathcal{M}.
\end{equation}
As both quantities has to be conserved during the process, we may conclude that the position of the torus center at $r = r_0$ stays unaffected. This is an artefact of the reflection symmetry with respect to $\bx=0$ that appears in the limit $\beta\rightarrow 0$. The centers of thicker tori that do not have this symmetry will be, in general, slowly drifting in time toward the central objects. 

\begin{figure}
  \begin{center}
    \FigureFile(0.45\textwidth,0.45\textwidth){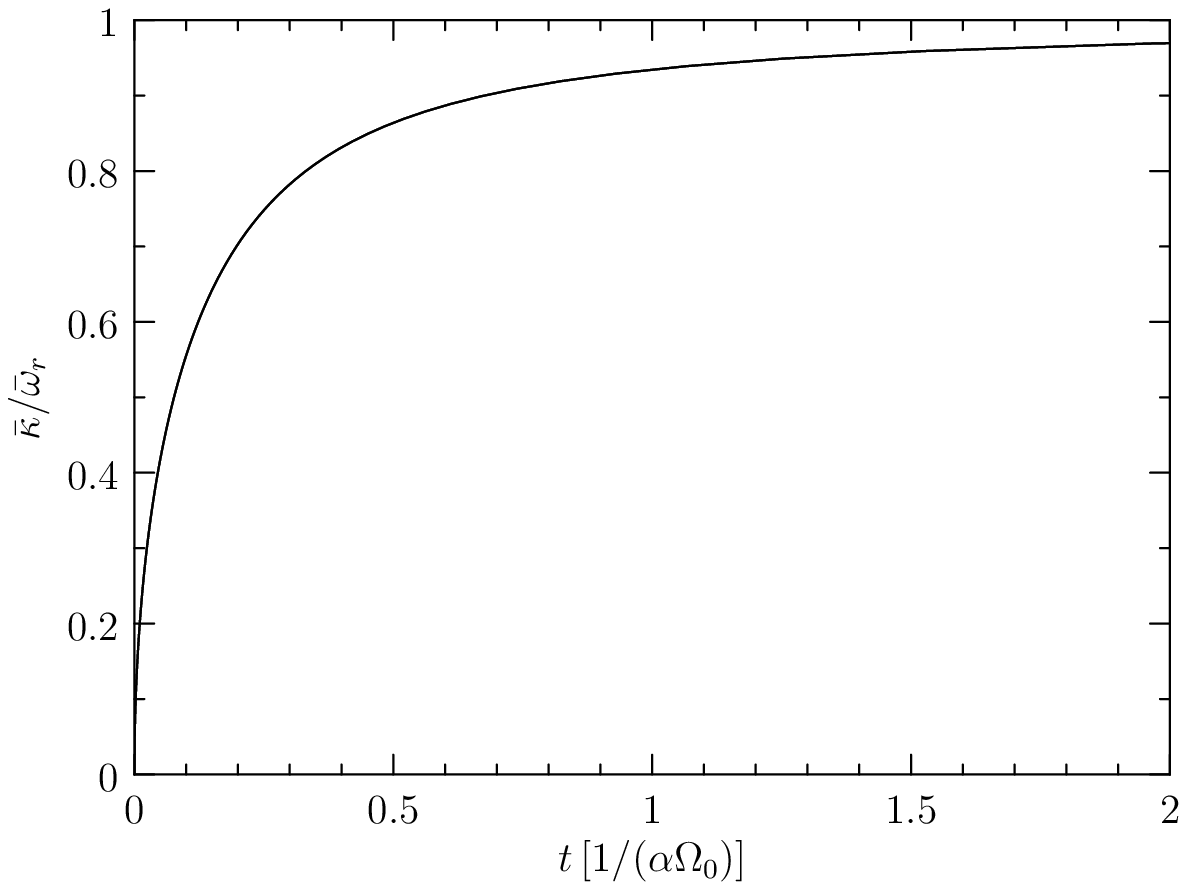} 
    \\
    \FigureFile(0.45\textwidth,0.45\textwidth){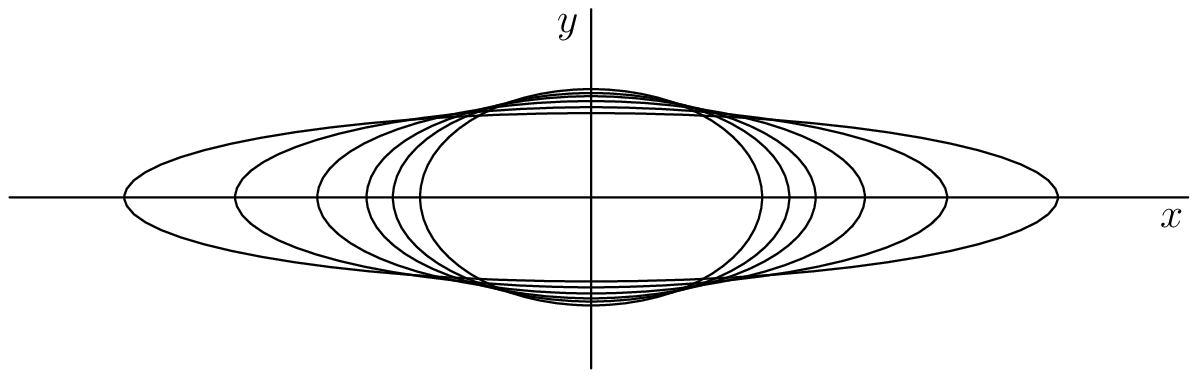} 
  \end{center}
  \caption{Top: A slow secular evolution of a slender torus from initially constant angular momentum ($\bar{\kappa}=0$) towards the Keplerian distribution ($\bar{\kappa}_r = \bar{\omega}_r$). In the calculation the value of the radial epicyclic frequency has been taken to correspond to the radius $r_0 = 10 M$ in Schwarzschild spacetime. Bottom: Shape of the torus at various times ($t\times\alpha \Omega_0=0$, 0.1, 0.2, 0.4, 0.8 and 1.5).}
  \label{fig:tori}
\end{figure}

Substituting the approximate expressions (\ref{eq:f}) for $f$ with $\bkappa=\bkappa(t)$ into equations (\ref{eq:ns-toroidal}) and (\ref{eq:viscevol-continuity}) and assuming a linear profile of $\ell(r)$, we obtain 
\begin{eqnarray}
  \frac{d\ln\beta^2}{dt} &+& \frac{d\bar{\kappa}^2}{dt}\bx^2 + \nonumber \\ 
  &+&\frac{f^{n-1}}{n \beta r_0}\left[\frac{\partial}{\partial\bx}(f^n v^r) + 
  \frac{\partial}{\partial\by}(f^n v^z)\right] = 0
\end{eqnarray}
and
\begin{equation}
  \left[\frac{d\bar{\kappa}^2}{dt} - \alpha\frac{\ell_0}{r_0^2}
  (4-\bar{\kappa}^2)(\bar{\omega}_r^2 - \bar{\kappa}^2)\right]\beta\bx + \frac{1}{r_0}\bar{\kappa}^2 v^r = 0.
\end{equation}
Now, making an ansatz 
\begin{equation}
  v^r(t,\bx,\by) = \beta \bx V^r(t), \quad v^z(t,\bx,\by) = \beta \by V^z(t),
\end{equation}
we obtain polynomial equations in $\bx$, $\by$. Comparing appropriate coefficients, we finally arrive at a set of four equations
\begin{eqnarray}
  n \frac{d\ln\beta^2}{dt} &=& -\frac{1}{r_0}(V^r + V^z), \\
  n \frac{d\bar{\kappa}^2}{dt} &=& \frac{1}{r_0}\left[(2n+1)V^r + V^z\right]\left(\bar{\omega}_r^2 - \bar{\kappa}^2\right), \\
  0 &=& \frac{1}{r_0}\left[V^r + (2n+1)V^z\right]\bar{\omega_z}^2, \\
  \frac{d\bar{\kappa}^2}{dt} &=& \frac{\alpha\ell_0}{r_0^2}\left(4-\bar{\kappa}^2\right)
  \left(\bar{\omega}_r^2 - \bar{\kappa}^2\right) - \frac{1}{r_0}\bar{\kappa}^2 V^r
\end{eqnarray}
for unknowns $\bkappa(t)$, $\beta(t)$, $V^r(t)$ and $V^z(t)$. The time of the secular evolution from the constant-angular momentum torus to the configuration characterized by $\bar{\kappa}$ is
\begin{eqnarray}
  t &=& \frac{1}{\alpha\Omega_0}\int_0^{\bar{\kappa}^2(t)}
  \frac{\bar{\omega}_r^2 - q k}{(4-k)(\bar{\omega}_r^2 - k)^2} dk 
  \nonumber \\
  &=& \frac{1}{\alpha\Omega_0(4-\bar{\omega}_r^2)^2}\Big[
  (4q -\bar{\omega}_r^2)\ln\frac{\bar{\omega}_r^2(4-\bar{\kappa}^2)}{4(\bar{\omega}_r^2 - \bar{\kappa}^2)}+
  \nonumber \\
  &~&(1-q)\bar{\kappa}^2\frac{4-\bar{\omega}_r^2}{\bar{\omega}_r^2 - \bar{\kappa}^2}\Big],
\end{eqnarray}
where $q = (2n+3)/(4n+4)$.

The solution $\bar{\kappa}(t)$ is shown in the top panel of Figure~\ref{fig:tori}. The angular momentum from the inner part of the torus is transported outward; its distribution becomes steeper approaching the Keplerian one (corresponding to $\bar{\kappa}=\bar{\omega}_r$) at long times.   A characteristic timescale of the viscous diffusion is 
\begin{equation}
  t_\mathrm{visc}=\alpha^{-1}\Omega_0^{-1}.
\end{equation}
The bottom panel shows the shape of the torus at several moments during the secular evolution. The torus becomes wider in a radial direction and shrinks in the vertical, as the angular momentum distribution approaches the Keplerian one. 

In the next sections we calculate damping rates of the lowest-order oscillation modes due to viscosity in tori with arbitrary angular-momentum distributions. Since the oscillations occur on timescales that are much shorter than $t_\mathrm{visc}$, we will ignore the viscous diffusion in the rest of the paper.

\section{Perturbations}
\label{sec:pert}
The unperturbed state is axisymmetric and can be regarded as stationary on timescales shorter than $t_\mathrm{visc}$. Therefore the $t$ and $\phi$-dependence of the perturbation is $\propto\exp[-\ii(\omega t - m\phi)]$, where $\omega$ is eigenfrequency of the oscillation mode and $m$ its azimuthal wavenumber. \citet{Papaloizou+Pringle1984} showed that it is convenient to express all perturbations in terms of a single quantity $W=-\delta p/(\rho\,\sigma)$, where $\sigma=\omega-m\Omega$. Linear Euler perturbation of the continuity equation gives
\begin{eqnarray}
  \frac{1}{r}\frac{\partial}{\partial r}(r f^n\delta v^r) + 
  \frac{\partial}{\partial z}(f^n\delta v^z) &+&
  \ii m f^n \delta v^\phi + 
  \nonumber\\
  &+& \ii\frac{2 n\bsigma^2}{\beta^2 r_0^2} f^{n-1}\, W = 0,
  \label{eq:continuity0}
\end{eqnarray}
where $\bsigma = \sigma/\Omega_0$. The perturbation of the velocity $\delta v^i$ can be calculated from $W$ using perturbed Navier-Stokes equations, which takes the form
\begin{eqnarray}
  \ii\,\delta v^r + \frac{2r\Omega}{\sigma}\delta v^\phi +
  \frac{\partial W}{\partial r} - \frac{m}{\sigma}\frac{d\Omega}{d r}W
  &=& -\frac{1}{\sigma}\delta \Fv^r,
  \label{eq:ns-r-l}
  \\
  \ii\,\delta v^\phi - \frac{\bar{\kappa}^2}{2r\Omega\sigma} \delta v^r+ 
  \frac{\ii m}{r^2} W 
  &=& -\frac{1}{\sigma} \delta \Fv^\phi,
  \label{eq:ns-phi-l}
  \\
  \ii\,\delta v^z + \frac{\partial W}{\partial z} &=& 
  -\frac{1}{\sigma} \delta \Fv^z.
  \label{eq:ns-z-l}
\end{eqnarray}
The perturbation of the viscous force in terms of the velocity perturbation $W$ is
\begin{eqnarray}
  \delta\Fv^r &=& \frac{\alpha\Omega_0 f^{-n}}{2(n+1)}\Big\{
  \frac{\partial}{\partial\by}\left[f^{n+1}\left(
  \frac{\partial\delta v^r}{\partial\by}+
  \frac{\partial\delta v^z}{\partial\bx}\right)\right] 
  \nonumber\\ &\phantom{=}&
  +\frac{\partial}{\partial\bx}\Big[f^{n+1}\Big(
  [2+a]\frac{\partial\delta v^r}{\partial\bx}+
  a\frac{\partial\delta v^z}{\partial\by} +
  \nonumber\\ &\phantom{=}&
  + a m\,\ii\delta v^\phi \Big)\Big]\Big\},
  \label{eq:ns-r-r}
  \\
  \delta\Fv^\phi &=& -\frac{\alpha\sigma}{\Omega_0r_0\beta}\frac{d\Omega}{dr}
  \frac{\partial W}{\partial\bx} 
  \nonumber\\ &\phantom{=}&
  +\frac{\alpha\Omega_0 f^{-n}}{2(n+1)}\Big\{
  \frac{\partial}{\partial\bx}\left[f^{n+1}\left(\frac{\partial\delta v^\phi}{\partial\bx}
  +\frac{\ii m}{r^2} \delta v^r\right)\right] 
  \nonumber\\ &\phantom{=}&
  +\frac{\partial}{\partial\by}\left[f^{n+1}\left(\frac{\partial\delta v^\phi}{\partial\by}
  +\frac{\ii m}{r^2} \delta v^z\right)\right]\Big\}, 
  \label{eq:ns-phi-r}
  \\
  \delta\Fv^z &=& \frac{\alpha\Omega_0 f^{-n}}{2(n+1)}\Big\{
  \frac{\partial}{\partial\bx}\left[f^{n+1}\left(
  \frac{\partial\delta v^r}{\partial\by}+
  \frac{\partial\delta v^z}{\partial\bx}\right)\right] 
  \nonumber\\ &\phantom{=}&
  +\frac{\partial}{\partial\by}\Big[f^{n+1}\Big(
  a\frac{\partial\delta v^r}{\partial\bx}+
  [2+a]\frac{\partial\delta v^z}{\partial\by}
  \nonumber\\ &\phantom{=}&
  + a m\,\ii\delta v^\phi \Big)\Big]\Big\},
  \label{eq:ns-z-r}
\end{eqnarray}
where we keep only leading-order terms in the limit $\beta\rightarrow 0$ (i.e.\ we neglect effects of the azimuthal curvature of the torus). The equations (\ref{eq:ns-r-l})--(\ref{eq:ns-z-l}) together with (\ref{eq:ns-r-r})--(\ref{eq:ns-z-r}) are then solved for poloidal velocity in terms of $W$. This procedure is easier if we assume that $\partial W/\partial\bx\gtrsim 1$, i.e.\ that the wavelength of a perturbation is at least comparable to the size of the torus. Then, up to the first order in $\alpha$, and in the lowest order in $\beta$, the solutions are
\begin{eqnarray}
  \delta v^r &=& \frac{\ii}{\beta r_0}\frac{\sigma^2}{\sigma^2-\kappa^2}\frac{\partial W}{\partial\bx} 
  \nonumber\\ &\phantom{=}&
  -\alpha\frac{\sigma^2}{\sigma^2-\kappa^2}\frac{f^{-n}}{2(n+1)\bsigma\beta r_0}
  \Big\{2\frac{\partial}{\partial\by}\left[f^{n+1}
  \frac{\partial^2 W}{\partial\bx\partial\by}\right]
  \nonumber\\ &\phantom{=}&
  +\frac{\partial}{\partial\bx}\left[f^{n+1}
  \left([2+a]\frac{\partial^2 W}{\partial\bx^2} + 
  a\frac{\partial^2 W}{\partial\by^2}\right)\right]  
  \nonumber\\ &\phantom{=}&
  - 8(n+1)f^n\left(1-\frac{\kappa^2}{4\Omega^2}\right)\frac{\partial W}{\partial\bx}
  \nonumber\\ &\phantom{=}&
  +\frac{\kappa^2}{\sigma^2-\kappa^2}\Big[(3+a)\frac{\partial}{\partial x}
  \left(f^{n+1}\frac{\partial^2 W}{\partial\bx^2}\right) + 
  \nonumber\\ &\phantom{=}&
  2\frac{\partial}{\partial\by}\left(f^{n+1}\frac{\partial^2 W}{\partial\bx\partial\by}\right)\Big]
  \Big\}
  \label{eq:vr}
\end{eqnarray}
and
\begin{eqnarray}  
  \delta v^z &=& \frac{\ii}{\beta r_0}\frac{\partial W}{\partial\by} - 
  \nonumber\\ &\phantom{=}&
  \alpha \frac{f^{-n}}{2(n+1)\bsigma\beta r_0}
  \Big\{2\frac{\partial}{\partial\bx}\left[f^{n+1}
  \frac{\partial^2 W}{\partial\bx\partial\by}\right]
  \nonumber\\ &\phantom{=}&
  +\frac{\partial}{\partial\by}\left[f^{n+1}
  \left(a\frac{\partial^2 W}{\partial\bx^2} + 
  [2+a]\frac{\partial^2 W}{\partial\by^2}\right)\right]
  \nonumber\\ &\phantom{=}&
  +\frac{\kappa^2}{\sigma^2-\kappa^2}\Big[\frac{\partial}{\partial x}
  \left(f^{n+1}\frac{\partial^2 W}{\partial\bx\partial\by}\right) + 
  \nonumber\\ &\phantom{=}&
  a\frac{\partial}{\partial\by}\left(f^{n+1}\frac{\partial^2 W}{\partial\bx^2}\right)\Big]
  \Big\}
  \label{eq:vz}
\end{eqnarray}

Substituting equations (\ref{eq:vr}) and (\ref{eq:vz}) into the continuity equation (\ref{eq:continuity0}) and keeping only dominant terms when $\beta\rightarrow0$, we obtain a single operator equation for $W$,
\begin{equation}
  \bar{\sigma}^4 W + \bar{\sigma}^2 \hat{B} W + \hat{C} W = \ii\alpha \hat{F}(\bsigma)W
  \label{eq:pp}
\end{equation}
with
\begin{eqnarray}
  \hat{B} &=& \frac{f^{1-n}}{2n}\left[
  \frac{\partial}{\partial\bx}\left(f^n\frac{\partial}{\partial\bx}\right) +
  \frac{\partial}{\partial\by}\left(f^n\frac{\partial}{\partial\by}\right)\right] - \bkappa^2
  \\
  \hat{C} &=& -\bkappa^2\frac{f^{1-n}}{2n}
  \frac{\partial}{\partial\by}\left(f^n\frac{\partial}{\partial\by}\right)
\end{eqnarray}
and
\begin{eqnarray}
  \hat{F}(\bsigma) &=& -\frac{\bsigma^2-\bkappa^2}{4n(n+1)\bsigma f^{n-1}}\Big\{
  \nonumber\\ &\phantom{=}& 
  a\left[\frac{\partial^2}{\partial\bx^2}
  \left(f^{n+1}\frac{\partial^2}{\partial\by^2}\right) +
  \frac{\partial^2}{\partial\by^2}
  \left(f^{n+1}\frac{\partial^2}{\partial\bx^2}\right)
  \right] 
  \nonumber\\ &\phantom{=}& 
  +(2+a)\Big[\frac{\partial^2}{\partial\bx^2}
  \left(f^{n+1}\frac{\partial^2}{\partial\bx^2}\right)
  \nonumber\\ &\phantom{=}& 
  +\frac{\partial^2}{\partial\by^2}
  \left(f^{n+1}\frac{\partial^2}{\partial\by^2}\right)
  \Big]+ 
  4\frac{\partial^2}{\partial\bx\partial\by}
  \left(f^{n+1}\frac{\partial^2}{\partial\bx\partial\by}\right)
  \nonumber\\ &\phantom{=}& 
  -2(n+1)(4-\bkappa^2)\frac{\partial}{\partial\bx}
  \left(f^{n}\frac{\partial}{\partial\bx}\right)+
  \nonumber\\ &\phantom{=}& 
  \frac{\bkappa^2}{\bsigma^2-\bkappa^2}\Big[
  (3+a)\frac{\partial^2}{\partial\bx^2}\left(f^{n+1}\frac{\partial^2}{\partial\bx^2}\right)
  \nonumber\\ &\phantom{=}& 
  +3\frac{\partial^2}{\partial\bx\partial\by}\left(f^{n+1}\frac{\partial^2}{\partial\bx\partial\by}\right)
  \nonumber\\ &\phantom{=}& 
  +a\frac{\partial^2}{\partial\by^2}\left(f^{n+1}\frac{\partial^2}{\partial\bx^2}\right)
  \Big]\Big\}.
  \label{eq:F}
\end{eqnarray}

When $\alpha=0$, equation (\ref{eq:pp}) is the second-order eigenvalue problem and describes oscillation modes of the inviscid slender torus. If, in addition, $\bkappa=0$ (constant angular momentum tori), the equation (\ref{eq:pp}) is reduced to the first-order eigenvalue problem for the operator $\hat{B}$. In that case the operator $\hat{B}$ is simply related to the operator  $\hat{L}^{(0)}$ introduced in \cite{Blaes+2006} by $\hat{B} = 2n\hat{L}^{(0)}$. \citet{Blaes1985} and \citet{Blaes+2006, Blaes+2007} showed that this operator is Hermitian with respect to the scalar product 
\begin{equation}
  \left\langle U, V\right\rangle = 
  \int U^{\ast} V f^{n-1} \dd\bx\,\dd\by,
  \label{eq:ScalarProduct}
\end{equation}
and demonstrated how this fact can be used in a perturbation approach to slightly non-slender tori. \citet{Blaes1985} used this theory to calculate growth-rate of the Papaloizou-Pringle instability and, more recently, the same approach has been employed in calculation of the eigenfrequency corrections to the epicyclic modes \citep{Blaes+2007}. 

A small viscous damping of the oscillation modes can be treated in a similar way. In addition, in this paper we extend the perturbation approach  of \citet{Blaes+2007} to the case of a general angular momentum distribution. In appendix~\ref{sec:PerturbationTheory} we formulate first-order perturbation theory for the eigenvalue problem (\ref{eq:pp}). Due to the viscosity term on the right-hand side of the equation (\ref{eq:pp}), the eigenfrequency of the $\nu$-th mode is slightly changed from the value $\sigma_\nu^{0}$ that corresponds to the inviscid flow to
\begin{equation}
  \bsigma_\nu = \bsigma^{(0)}_\nu + \alpha\,\bsigma^{(1)}_\nu + \dots,
\end{equation}
The first-order correction is given by
\begin{equation}
  \bsigma_\nu^{(1)} = \frac{\ii}{2\bsigma_\nu^{(0)}}
  \frac{\left\langle W_\nu^{(0)}, \hat{F} W_\nu^{(0)}\right\rangle}
  {\left\langle W_\nu^{(0)}, \left(\hat{B} + 2\sigma_\nu^{(0)2}\right) W_\nu^{(0)}\right\rangle} 
  \label{eq:first-order-correction}
\end{equation}
with $W_\nu^{(0)}$ being the eigenfunction corresponding to the $\nu$-th mode of the inviscid torus.

Since the oscillations of the inviscid tori are not damped, the growth-rate of the viscous torus is given by
\begin{equation}
  \gamma_\nu = \alpha\Omega_0 \, \mathrm{Im}[\bsigma^{(1)}_\nu].
  \label{eq:gamma}
\end{equation}
The damping corresponds to the negative values of $\gamma$. 

\section{Special cases}
\label{sec:results-special}
\subsection{The WKBJ limit}
In the limit of short wave-lengths of the perturbation, the eigenfunctions can be approximated using WKBJ ansatz
\begin{equation}
  W\propto \exp\left[\ii\int k_x d\bx + \ii \int k_y d\by \right],
\end{equation}
where $k_x$ and $k_y$ are the horizontal and vertical components of the wavevector. \citet{Blaes+2006} identified two distinct classes of modes, acoustic and inertial that obey the dispersion relations
\begin{equation}
  \bsigma^2 = \frac{f}{2n} k^2
  \quad \mathrm{and} \quad
  \bsigma^2 = \bkappa^2 \frac{k_y^2}{k^2},
\end{equation}
respectively. While the frequency of the acoustic modes grows with increasing $k = (k_x^2 + k_y^2)^{1/2}$, that of the inertial modes is always smaller or at most equal $\bkappa$. 

After calculations briefly summarized in Appendix~\ref{sec:appendix-wkbj}, we find that the approximate expression for the growth rates of the acoustic modes is
\begin{equation}
  \gamma = -\frac{n}{n+1}\left(1+\frac{a}{2}\right)\bsigma^2\alpha\Omega_0,
\end{equation}
while that of the inertial modes is proportional to
\begin{equation}
  \gamma \sim \left(1 - \frac{\bsigma^2}{\bkappa^2}\right)\frac{\alpha\Omega_0}{(n+1)\lambda_\ast^2},
\end{equation}
where $\lambda_\ast$ is a typical wavelength of the mode. Therefore, while the high-frequency acoustic modes are always damped by the viscosity, the low-frequency inertial modes become unstable when the $\alpha$-viscosity is introduced.

\subsection{Keplerian angular momentum distribution}
In the limit of the Keplerian angular momentum distribution, the characteristic frequency of the inertial oscillations is equal to the radial epicyclic frequency, $\bkappa = \bomega_r$. The Lane-Embden function does not depend on $\bx$ and is a function of $\by$ only,
\begin{equation}
  f = 1 - \bomega_z^2\by^2.
\end{equation}
One set of the oscillation modes that depends linearly on $\bx$ have eigenfunctions given by the Gegenbauer polynomials $C_j^k$,
\begin{equation}
  W_j = C_j^{n-1/2}(\bomega_z\by),
  \quad\mathrm{or}\quad
  W_j = \bx C_j^{n-1/2}(\bomega_z\by),
\end{equation}
where $j\geq 1$ denotes a number of vertical nodes \citep[see][]{Blaes+2006}. The corresponding eigenfrequencies are
\begin{equation}
  \bsigma^2_j = \frac{1}{2n}\,j (j+2n-1)\bomega_z^2.
\end{equation}
The vertical epicyclic mode corresponds to $j=1$. Applying our theory we find that the growth rate of the $j$-th mode is
\begin{equation}
  \gamma_j = -\left(1+\frac{a}{2}\right)\frac{(j+2n)(j-1)}{2(n+1)}\bomega_z^2\alpha\Omega_0
\end{equation}
(see Appendix~\ref{sec:appendix-Kepler}). Hence, all the modes are damped in the Keplerian limit. In addition, for high-order modes with $j\gg 1$ the damping rate agrees with that predicted by the WKBJ approximation.

\section{Results for the lowest-order modes}
\label{sec:results}
\begin{figure}
  \begin{center}
    \FigureFile(0.45\textwidth,0.45\textwidth){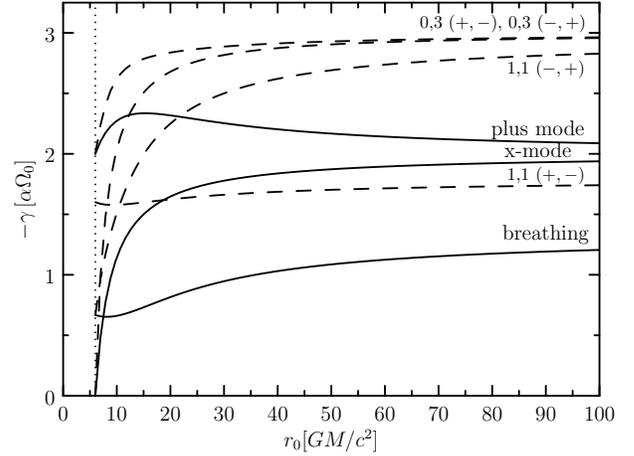}
  \end{center}
  \caption{Damping rates of the lowest-order modes of constant-angular momentum tori. The solid and dashed lines correspond to the second-order and third order modes, respectively. The third-order modes are labeled according to their $j$ and $k$ numbers and $x$ and $y$ parity \citep[see][]{Blaes+2006}.}
  \label{fig:rates}
\end{figure}

\citet{Blaes+2006} derived eigenfunctions and eigenfrequencies of the lowest-order modes of slender tori with arbitrary angular momentum distribution. The eigenfunctions of these modes are given by the polynomials of low-order in $\bx$ and $\by$. The most simple is the corotation mode corresponding to $\bsigma=0$ and a zero-order polynomial $W=1$. It may become unstable when $\beta>0$ and in that case it corresponds to the prinicipal mode of the Papaloizou-Pringle instability. The eigenfunctions linear in $\bx$ and $\by$ characterize two epicyclic modes -- in slender tori they corresponds to a rigid motion in radial or vertical direction with associated  epicyclic frequencies. Four second order modes (the X-mode, inertial mode, breathing mode and plus-mode) have eigenfunctions quadratic in $\bx$ and $\by$. Their velocity patterns have single node at the center of the torus. \citet{Blaes+2006} show how to construct modes of arbitrary order by solving appropriate sets of linear equations. In this section, we calculate viscous growth-rates of the lowest order modes (up to the third order) using the equations (\ref{eq:first-order-correction}) and (\ref{eq:gamma}). For this purpose we use formulae for the eigenfunctions and eigenfrequencies derived by \citet{Blaes+2006}.

\subsection{The epicyclic modes}
The radial and vertical epicyclic modes correspond to the linear eigenfunctions, $W_\mathrm{rad}=\bx$ or $W_\mathrm{vert}=\by$. In the inviscid slender torus, they describe uniform displacements from the equilibrium positions with frequencies equal to epicyclic frequencies of a freely moving particle. Therefore, the shear of the corresponding velocity field vanishes and so may vanish the viscous damping. In fact, this is the case of the vertical epicyclic mode for which $\hat{F} W_\mathrm{vert}=0$ and therefore $\gamma_\mathrm{vert} = 0$. In the case of the radial epicyclic modes, however, we find that
\begin{equation}
  \hat{F}W_\mathrm{rad} = \frac{1}{\bomega_r}(\bomega_r^2-\bkappa^2)(4-\bkappa^2)\bx
\end{equation}
and therefore the viscous growth-rate is
\begin{equation}
  \gamma_\mathrm{rad} = -\frac{1}{2}\left(1-\frac{\bkappa^2}{\bomega_r^2}\right)^2(4-\bkappa^2)\,\alpha\Omega_0.
  \label{eq:gamma-rad}
\end{equation}
This is because during the radial oscillations, the viscous force does not vanish even though the shear of the perturbed velocity field does. Because of variations in the kinematic viscosity coefficient $\eta$ (that arise due to a pressure variations), a nonzero azimuthal viscous force appears,
\begin{equation}
  \delta\mathcal{F}^\phi_\mathrm{visc} \propto -\frac{\alpha\omega_r}{2\beta r_0^2}(4-\bkappa^2)\exp[-\ii(\omega t - m\phi)].
\end{equation}
This force is in the anti-phase with the velocity perturbations
\begin{equation}
  \delta v^\phi \propto \frac{\bomega_r\bkappa^2}{2r(\bomega_r^2-\bkappa^2)\beta r_0}\exp[-\ii(\omega t - m\phi)],
\end{equation}
and therefore causes the damping of oscillations. Similar effect is absent in the case of the vertical oscillations because the $z\phi$-component of the  shear of the equilibrium velocity vanishes.

Equation (\ref{eq:gamma-rad}) shows that the growth-rate vanishes in the limit of the Keplerian angular momentum distribution. This is because the amplitude of the pressure variations decays when approaching that limit (note that $\delta p/p \sim (1-\bomega_r^2/\bkappa^2)(\delta v^\phi/\Omega)$ for the radial epicyclic mode). This is consistent with the behavior of $p$-modes in geometrically thin accretion disks. Although their amplitudes grow, the growth-rates are inversely proportional to the squared wavelengths of the modes \citep[e.g.][]{Kato1978, Kato+1998}. The $p$-modes with infinite wavelength (that corresponds to our radial epicyclic mode) are not affected by the viscosity.

\subsection{Constant-angular-momentum tori}

\begin{figure}
  \begin{center}
    \FigureFile(0.45\textwidth,0.45\textwidth){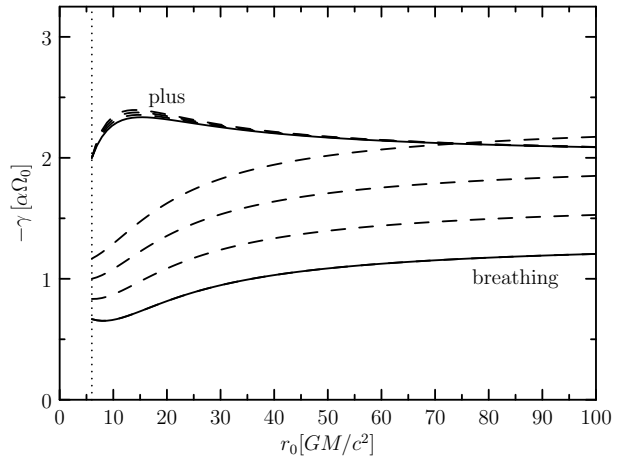}
  \end{center}
  \caption{Damping rates of the plus mode and the breathing mode for different values of the bulk viscosity. The solid lines correspond to the case $a=-2/3$ ($\xi=0$), the dashed lines are for $a=-1/3, 0$ and $1/3$ ($\xi = \eta/3, 2\eta/3$ and $\eta$, respectively).}
  \label{fig:damprates-a}
\end{figure}

Figure~\ref{fig:rates} shows a radial dependence of damping rates (in units of $\alpha\Omega_0$) of the second-order (solid line) and third order modes (dashed lines). The third-order modes have been labeled according to the convention of \citet{Blaes+2006}. The coefficient $a$ was set to $a=-2/3$, what corresponds to vanishing bulk viscosity coefficient, $\xi=0$. One may notice that the gravitational field enters into all formulae only through the expressions for the orbital and epicyclic frequencies $\Omega_0$, $\bomega_r$ and $\bomega_z$. Similarly to other authors, we use the Schwarzschild expressions for these frequencies, despite the fact that the formulae have been derived in the framework of Newtonian physics. We believe that this inconsistency will not affect the qualitative discussion of the paper significantly. The radial coordinate of the torus center is shown in units of the gravitational radii $GM/c^2$.

The analytic formulae for damping rates are rather complicated. The exception is the X-mode, whose growth/damping rate is
\begin{equation}
  \gamma_\times = 
  -\frac{2\,\bomega_r^2\left(1+\bomega_z^2\right)}{\bomega_r^2 + \bomega_z^2}\,\alpha\Omega_0.
\end{equation}
The damping rate does not depend on the bulk viscosity coefficient $a$ because the X-mode oscillations are incompressible. On the other hand, both plus and breathing modes weakly depend on the value of $a$ as it is shown in Figure~\ref{fig:damprates-a}.

Figure~\ref{fig:rates} suggests that the the ratios $\gamma_\nu/(\alpha\Omega_0)$ tend to the constant for $r\gg GM/c^2$. This behavior nicely illustrate the scaling invariance of the Newtonian gravity. The Newtonian value of the damping rates of the plus and the breathing modes are $\gamma_{+}=2\alpha\Omega_0$ and $\gamma_\mathrm{breath}=(2+a)\alpha\Omega_0$.

\subsection{Dependence on the angular-momentum profile}

\begin{figure}
  \begin{center}
    \FigureFile(0.45\textwidth,0.45\textwidth){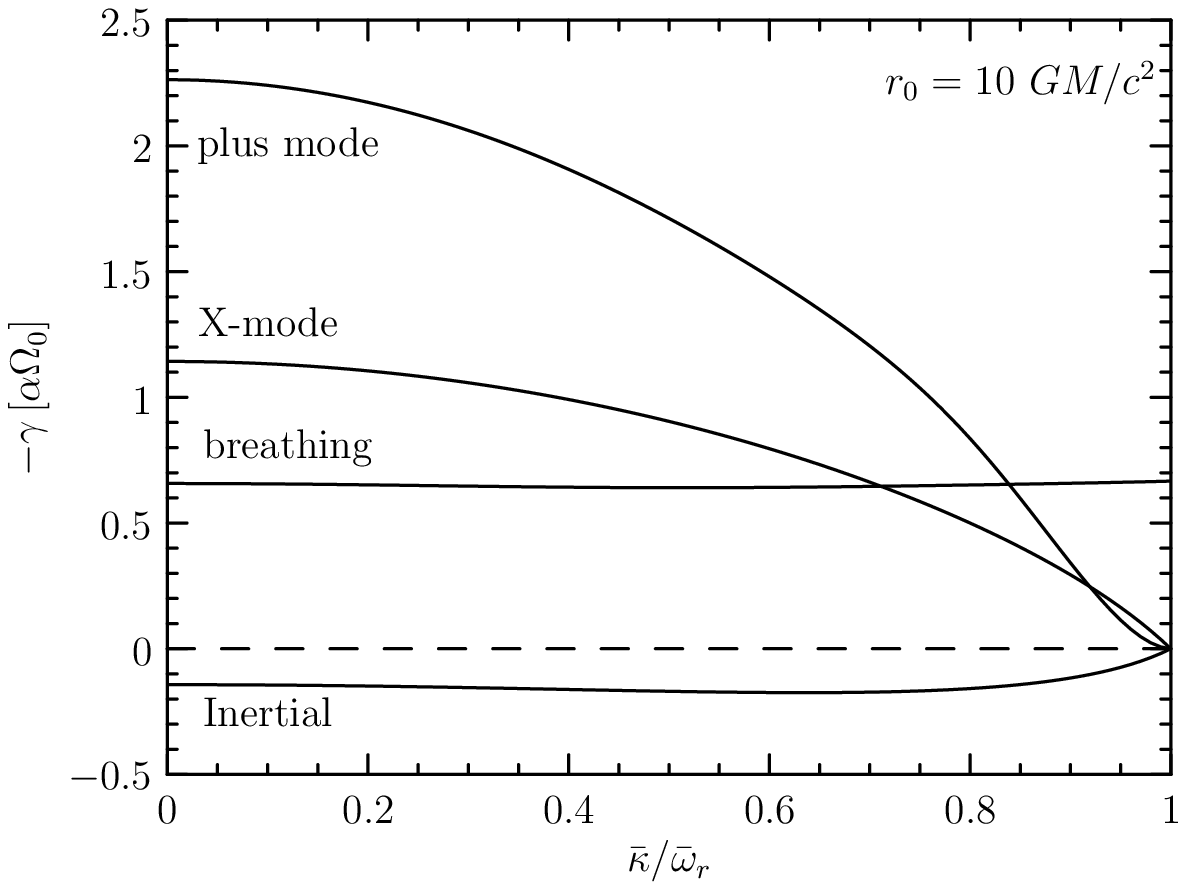}
    \\
    \FigureFile(0.45\textwidth,0.45\textwidth){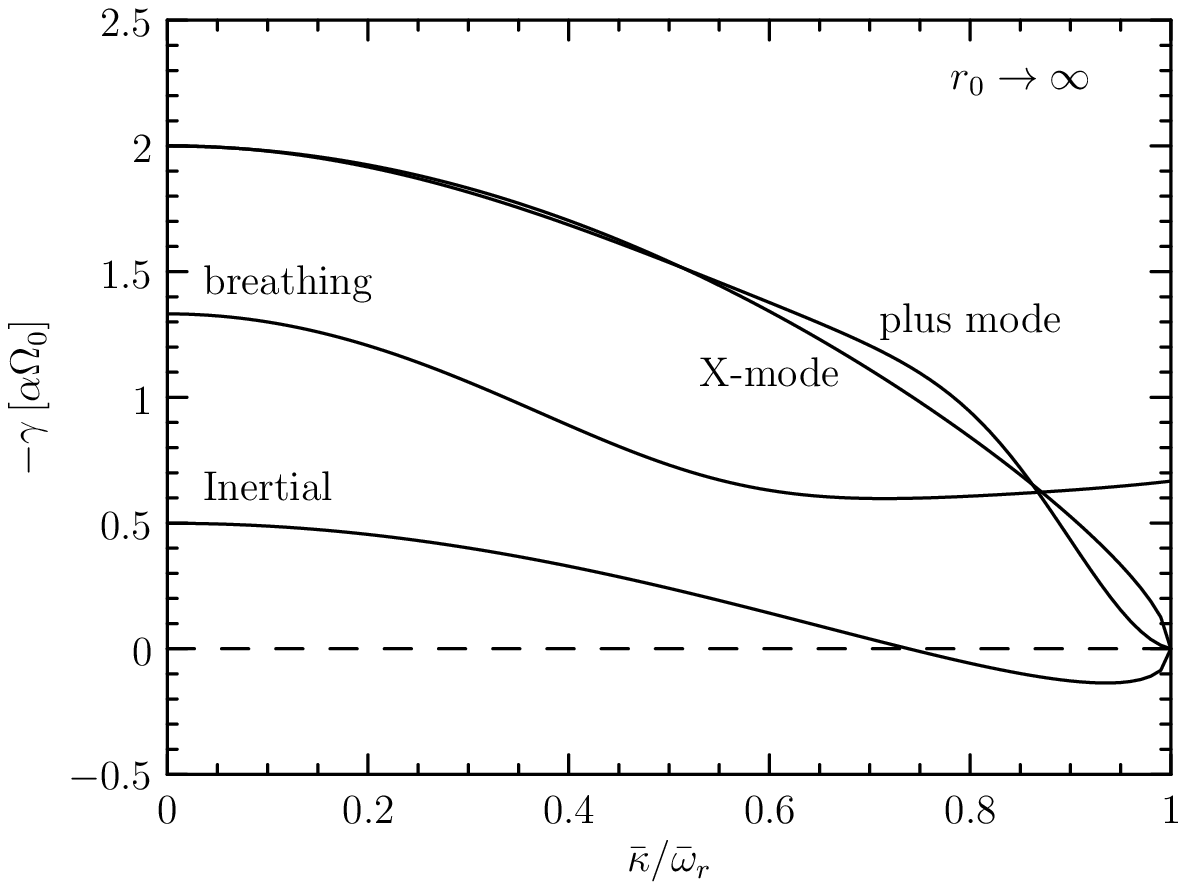}
  \end{center}
  \caption{Damping rates of the lowest-order modes as functions of the angular momentum distribution. The top panel corresponds to the torus placed at $r_0=10 GM/c^2$, the bottom panel shows Newtonian limit $r\rightarrow\infty$. In both situations we set $a=-2/3$ what corresponds to vanishing bulk viscosity coefficient.}
  \label{fig:rates-l}
\end{figure}

The behavior of the damping rates of the four lowest order modes with changing angular momentum distribution is shown in Figure~\ref{fig:rates-l}. The top panel is devoted to the situation when torus is placed close to the central object at radius $r_0=10GM/c^2$. The lower panel shows the same for the torus at large radii and corresponds to the Newtonian limit. The damping rates of all the modes except the breathing one are significantly reduced for Keplerian angular-momentum distribution. Although our analysis predicts that the damping of all three modes vanishes completely, it is still possible that the actual damping rate is of a higher order in $\beta$, say on the timescale of $\gtrsim(\beta\alpha\Omega)^{-1}$.

The growth-rate of the breathing mode in the Keplerian limit ($\bkappa\rightarrow 1$) is
\begin{equation}
  \gamma_\mathrm{breath} = -\bomega_z^2\left(1+\frac{a}{2}\right)\alpha\Omega_0.
\end{equation}

\begin{figure}
  \begin{center}
    \FigureFile(0.45\textwidth,0.45\textwidth){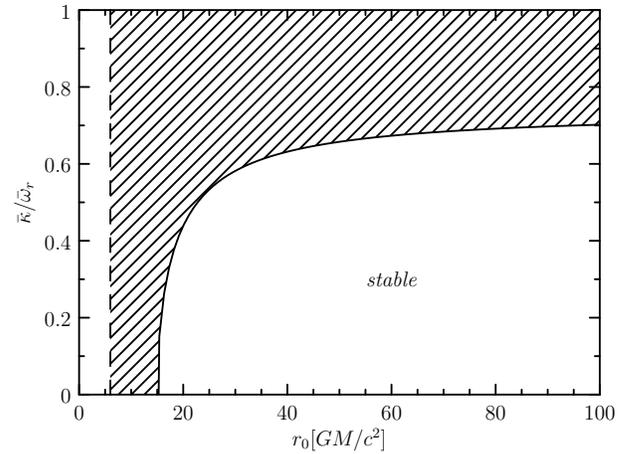}
  \end{center}
  \caption{Instability threshold for the inertial mode.}
  \label{fig:treshold}
\end{figure}

Our results also suggest that the torus is unstable with respect to the inertial mode. In the relativistic case (top panel) this mode grows in time independently on $\bkappa$, while in the Newtonian limit this growth is suppressed for less steep angular momentum profiles. Figure~\ref{fig:treshold} shows regions of stable and unstable configurations in the $(r,\bkappa)$-plane.

\section{Discussion and conclusions}
\label{sec:concl}
In this paper we have studied viscosity effects in slender tori. Due to the angular-momentum transport, the torus becomes wider in the horizontal direction and thinner in the vertical as the angular momentum distribution approach the Keplerian one. We have shown that this process occurs on the timescale $\sim\alpha^{-1}\Omega_0^{-1}$ and can be regarded as a quasi-steady, i.e.\ as a series of steady inviscid tori configurations, because the poloidal velocity due to the viscosity is negligible. We have also found that the angular momentum transport is symmetric, the location of the center of the torus remains unchanged. This result is related to the reflection symmetry around the $\bx=0$ surface that appears in the slender torus limit. Including the higher-order terms in $\beta$ that break this symmetry, we would likely find a shift of the torus center towards the central object. 

In the rest of the paper we dealt with the oscillations of the viscous tori. For simplicity, we neglected the secular evolution what allowed as to assume the harmonic time dependence of perturbations. We derived inhomogeneous master equation (\ref{eq:pp}) with left-hand side giving the normal modes of the inviscid tori and right-hand side describing effects of viscosity on the modes. We formulated a general first-order perturbation theory for the case when the right-hand side is small and with the aid of this theory we obtained a general formula for the mode growth rates. Applying this formula in the short-wave limit, we have found that the viscosity acts to damp the acoustic modes. On the other-hand the viscous tori are unstable with respect to the inertial oscillations. The growth-rate of the instability increases with decreasing wavelength. We also calculated the damping rates of the acoustic oscillations in the Keplerian limit and showed that they agree with the results based on the WKBJ theory in the limit of high nodal numbers. Finally we calculated the growth rates of the lowest order modes and explored their behavior with changing various parameters. We found that the damping of three of four modes is significantly reduced as the angular momentum distribution approaches the Keplerian one. 

\medskip
We gratefully acknowledge kind hospitality of Mrs. Malwina from Stary Giera{\l}t{\'o}w, where the work on this project was initiated. JH also acknowledges many helpful discussions with his colleagues from Prague Astronomical Institute and support of the Grant Agency of the Czech Republic (project no. P209/11/2004). MAA acknowledges the support of the Czech grant MSM 4781305903.


\appendix

\section{First-order perturbation theory for tori with arbitrary 
angular-momentum distribution}
\label{sec:PerturbationTheory}
Here we deal with a perturbation theory for equation (\ref{eq:pp}) with the assumption that $\alpha$ is small. The right-hand side is therefore treated as a perturbation. We start with the observation that both, the operators $\hat{B}$ and $\hat{C}$ are Hermitian with respect to the scalar product (\ref{eq:ScalarProduct}), i.e. \
\begin{eqnarray}
  \left\langle \hat{B} W_\mu^{(0)}, W_\nu^{(0)}\right\rangle &=&
  \left\langle W_\mu^{(0)}, \hat{B} W_\nu^{(0)}\right\rangle \\
  \left\langle \hat{C} W_\mu^{(0)}, W_\nu^{(0)}\right\rangle &=&
  \left\langle W_\mu^{(0)}, \hat{C} W_\nu^{(0)}\right\rangle.
\end{eqnarray}

The eigenfunctions $W_\nu^{(0)}$ that solve equation (\ref{eq:pp}) with $\alpha=0$ form a complete set and can be therefore used as a basis with respect to which solutions of the perturbed equation will be expanded. However, contrary to the first-order eigenvalue problems, the eigenfunctions  $W_\nu^{(0)}$ are not orthogonal. Instead, they satisfy so the pseudo-orthogonality relation
\begin{eqnarray}
  -\left(\bsigma_\mu^{(0)2} + \bsigma_\nu^{(0)2}\right)
  \left\langle W_\mu^{(0)}, W_\nu^{(0)}\right\rangle + 
  \nonumber \\
  \left\langle W_\mu^{(0)}, \hat{B} W_\nu^{(0)}\right\rangle = 0
\end{eqnarray}
for $\mu\neq\nu$.

The solutions of the perturbed problem are then expanded via $\alpha$ as
\begin{eqnarray}
  \bsigma_\nu &=& \bsigma^{(0)}_\nu + \alpha\,\bsigma^{(1)}_\nu + \dots,
  \\
  W_\nu &=& W^{(0)}_\nu + \alpha \sum_\mu c_{\nu\mu} W^{(0)}_\mu + \dots
\end{eqnarray}
and substituted into equation (\ref{eq:pp}). Comparing coefficient of the same powers of $\alpha$ we obtain in the first order
\begin{eqnarray}
  -\sum_\mu c_{\nu\mu} \left(\bsigma_\mu^{(0)2} - \bsigma_\nu^{(0)2}\right)
  \left[\left(\bsigma_\mu^{(0)2} + \bsigma_\nu^{(0)2}\right)+\hat{B}\right]W_\mu^{0} 
  \nonumber \\
  +2\bsigma_\nu^{(0)}\bsigma_\nu^{(1)}\left[2\bsigma_\nu^{(0)2} + 
  \hat{B}\right]W_\nu^{(0)} = \ii \hat{F} W_\nu^{(0)}.
\end{eqnarray}
If we now perform scalar product with $W_\nu^{(0)}$ and use the pseudo-orthogonality relation, we get
\begin{eqnarray}
  2\bsigma_\nu^{(0)}\bsigma_\nu^{(1)}\left\langle W_\nu^{(0)}, 
  \left(\hat{B} + 2\sigma_\nu^{(0)2}\right) W_\nu^{(0)}\right\rangle = 
  \nonumber \\
  \ii\left\langle W_\nu^{(0)}, \hat{F} W_\nu^{(0)}\right\rangle,
\end{eqnarray}
from which we imediatelly recover the relation (\ref{eq:first-order-correction}).

\section{The WKBJ approximation}
\label{sec:appendix-wkbj}
In the WKBJ limit, $k\rightarrow\infty$, we find that
\begin{eqnarray}
  \hat{F}W &=& -\frac{(\bsigma^2-\bkappa^2)f^2 k^2}{4n(n+1)} \times
  \nonumber \\ &\times&
  \left[(2+a)k^2 + \frac{(3+a)\bkappa^2}{\bsigma^2-\bkappa^2}k_x^2\right] W,
  \label{eq:wkbj-fw}
\end{eqnarray}
and
\begin{equation}
  (\hat{B}+2\bsigma^2) W = \left[-\frac{1}{2n}f k^2 + 2\bsigma^2 - \bkappa^2\right]W.
  \label{eq:wkbj-bw}
\end{equation}

For the acoustic modes, the second term in the square bracket of equation (\ref{eq:wkbj-fw}) vanishes because $\bsigma\gg\bkappa$ and using the corresponding dispersion relation, we obtain
\begin{equation}
  \left\langle W,\hat{F}W\right\rangle = -\frac{(2+a)n\bsigma^5}{n+1}\left\langle W,W\right\rangle.
\end{equation}
Similarly, we find that
\begin{equation}
  \left\langle W,(\hat{B} + 2\bsigma^2)W\right\rangle = \bsigma^2\left\langle W,W\right\rangle.
\end{equation}
With aid of the equations (\ref{eq:first-order-correction}) and (\ref{eq:gamma}), we therefore find that the growth/damping rate is
\begin{equation}
  \gamma = -\frac{n}{n+1}\left(1 + \frac{a}{2}\right)\bsigma^2 \alpha\Omega_0.
\end{equation}

In the case of the inertial modes we find that
\begin{equation}
  \left\langle W,\hat{F}W\right\rangle = -\frac{\bkappa^2 - \bsigma^2}{4n(n+1)\bsigma}\left\langle W, k^4f^2 W \right\rangle
\end{equation}
and
\begin{equation}
  \left\langle W,(\hat{B} + 2\bsigma^2)W\right\rangle = -\frac{1}{2n}\left\langle W, k^2 f W \right\rangle.
\end{equation}
The growth rate of these modes is therefore
\begin{equation}
  \gamma = \frac{\bkappa^2 - \bsigma^2}{2(n+1)\bsigma^2}
  \frac{\left\langle W, k^4f^2 W \right\rangle}{\left\langle W, k^2 f W \right\rangle}\alpha\Omega_0.
\end{equation}
Both scalar products on the right-hand side are positive as their are integrals of the positive quantities. Therefore they cannot change the sign of $\gamma$ and we can conclude that the high-order inertial modes grow. Moreover, introducing a typical wavelength of the mode $\lambda_\ast$, the upper and the lower scalar product can be estimated as $\sim\lambda_\ast^{-4}\langle W,W \rangle$ and $\sim\lambda_\ast^{-2}\langle W,W \rangle$, respectively. Hence the growth rate can be estimated as
\begin{equation}
  \gamma \sim \left(1 - \frac{\bsigma^2}{\bkappa^2}\right)\frac{\alpha\Omega_0}{(n+1)\lambda_\ast^2}.
\end{equation}

\section{Damping of the oscillations in the Keplerian limit}
\label{sec:appendix-Kepler}
In the Keplerian limit, the Lane-Embden function $f$ is independent of $\bx$ and the eigenfunctions $W_j$ depend on $\bx$ at most linearly. Applying the operator $\hat{F}$ given by relation (\ref{eq:F}) on $W$ we obtain
\begin{equation}
  \hat{F}W_j = -\frac{(\bsigma^2-\bomega_r^2)(2+a)}{4n(n+1)\bsigma_j f^{n-1}} \frac{\partial^2}{\partial\by^2}
  \left(f^{n+1}\frac{\partial^2 W_j}{\partial\by^2}\right).
\end{equation}
Performing twice the integration by parts in $\by$, we find that the scalar product $\langle W_j, \hat{F} W_j\rangle$ is given by
\begin{equation}
  \langle W_j, \hat{F} W_j\rangle \sim -\frac{(\bsigma^2-\bomega_r^2)(2+a)}{4n(n+1)\bsigma} I^2,
  \label{eq:appendix-Kepler-numerator}
\end{equation}
where  
\begin{equation}  
  I^2 = \int\left|\frac{\partial^2 W_j}{\partial\by^2}\right|^2 f^{n+1}(\by) d\by,
\end{equation}
where we omitted the factor coming from the integration over $\bx$. Similarly,
\begin{equation}
  \left\langle W_j,(\hat{B}+2\bsigma_j^2)W_j\right\rangle = -\frac{1}{2n}I^1 + (2\bsigma_j^2-\bkappa_j^2)I^0,
  \label{eq:appendix-Kepler-denominator}
\end{equation}
where
\begin{equation}  
  I^0 = \int\left|W_j\right|^2 f^{n-1}(\by) d\by,
\end{equation}
and
\begin{equation}  
  I^1 = \int\left|\frac{\partial W_j}{\partial\by}\right|^2 f^{n}(\by) d\by.
\end{equation}
Using the substitution $\zeta = \bomega_z\by$ and with the aid of standard rules for derivatives of the Gegenbauer polynomials and their normalization, the integrals $I^0$, $I^1$ and $I^2$ can be evaluated as
\begin{eqnarray}
  I^0 &=& \frac{1}{\bomega_z}\int_{1}^1\left[C_j^{n-1/2}\right]^2\left(1-\zeta^2\right)^{n-1}d\zeta
  \nonumber \\
  &=& \frac{\pi\Gamma(j+2n-1)}{\bomega_z 4^{n-1}(j+n+1/2)\Gamma^2(n-1/2)\Gamma(j+1)},
\end{eqnarray}
\begin{eqnarray}
  I^1 &=& \bomega_z^3\int_{-1}^1\left[\frac{d}{d\zeta} C_j^{n-1/2}\right]^2 \left(1-\zeta^2\right)^{n}d\zeta 
  \nonumber \\
  &=&(2n-1)^2\bomega_z\int_{-1}^1\left[C_{j-1}^{n+1/2}\right]^2\left(1-\zeta^2\right)^{n}d\zeta  
  \nonumber \\
  &=& -\frac{\pi(2n-1)^2\bomega_z\Gamma(2n+j)}{4^{n}(j+n-1/2)\Gamma^2(n+1/2)\Gamma(j)}.
\end{eqnarray}
and
\begin{eqnarray}
  I^2 &=& \bomega_z^3\int_{-1}^1\left[\frac{d^2}{d\zeta^2} C_j^{n-1/2}\right]^2 \left(1-\zeta^2\right)^{n+1}d\zeta 
  \nonumber \\
  &=&(2n-1)^2(2n+1)^2\bomega_z^3\int_{-1}^1\left[C_{j-2}^{n+3/2}\right]^2\left(1-\zeta^2\right)^{n+1}d\zeta  
  \nonumber \\
  &=& -\frac{\pi(2n-1)^2(2n+1)^2\bomega_z^3\Gamma(j+2n+1)}{4^{n+1}(j+n-1/2)\Gamma^2(n+3/2)\Gamma(j-1)}.
\end{eqnarray}
Then, substituting into equations (\ref{eq:appendix-Kepler-numerator}) and (\ref{eq:appendix-Kepler-denominator}) and then finally into the equations (\ref{eq:first-order-correction}) and (\ref{eq:gamma}), we obtain
\begin{equation}
 \gamma_j = -\left(1+\frac{a}{2}\right)\frac{(j+2n)(j-1)}{2(n+1)}\bomega_z^2\alpha\Omega_0.
\end{equation}


\begin{thebibliography}{}
\bibitem[{{Blaes}(1985)}]{Blaes1985}
{Blaes} O.M., 1985, \mnras, 216, 553

\bibitem[{{Blaes} et~al.(2006){Blaes}, {Arras}, \& {Fragile}}]{Blaes+2006}
{Blaes} O.M., {Arras} P., {Fragile} P.C., 2006, \mnras, 369, 1235

\bibitem[{{Blaes} et~al.(2007){Blaes}, {{\v S}r{\'a}mkov{\'a}}, {Abramowicz},
  {Klu{\'z}niak}, \& {Torkelsson}}]{Blaes+2007}
{Blaes} O.M., {{\v S}r{\'a}mkov{\'a}} E., {Abramowicz} M.A., {Klu{\'z}niak} W.,
  {Torkelsson} U., 2007, \apj, 665, 642

\bibitem[{{Blumenthal} et~al.(1984)}]{Blumenthal+1984} Blumenthal, G.~R., 
Lin, D.~N.~C., \& Yang, L.~T.\ 1984, \apj, 287, 774 

\bibitem[{{Kato}(1978)}]{Kato1978}
{Kato} S., \mnras, 185, 629 

\bibitem[Kato(1991)]{Kato1991} 
Kato, S.\ 1991, \pasj, 43, 557 

\bibitem[Kato(1994)]{Kato1994}
Kato, S.\ 1994, \pasj, 46, 415 

\bibitem[Kato et al.(1998)]{Kato+1998} 
Kato, S., Fukue, J., \& Mineshige, S.\ 1998, 
Black-hole accretion disks. Kyoto University Press, 1998.

\bibitem[Nowak \& Wagoner(1992)]{Nowak-Wagoner1992} 
Nowak, M.~A., \& Wagoner, R.~V.\ 1992, \apj, 393, 697 

\bibitem[{{Papaloizou} \& {Pringle}(1984)}]{Papaloizou+Pringle1984}
{Papaloizou} J.C.B., {Pringle} J.E., 1984, \mnras, 208, 721

\bibitem[Shakura \& Sunyaev(1973)]{Shakura-Sunyaev1973} 
Shakura, N.~I., \& Sunyaev, R.~A.\ 1973, \aap, 24, 337 


\end{thebibliography}
\end{document}